# Growth of bioluminescent bacteria under modelled gravity of different astronomical bodies in the Solar system

Santosh Bhaskaran*[1,@], Rohan Dudhale[1,2,$], Jyotsana Dixit[1,2], Ajit Sahasrabuddhe[1,3,#] and Pandit B. Vidyasagar[1]

[1]Biophysics Laboratory, Department of Physics, University of Pune, Pune 411 007, INDIA

[2]Department of Microbiology, P.E.S. Modern College of Arts, Science and Commerce, Ganeshkhind, Pune 411 053 INDIA. [$] Haffkine Institute for Training, Research and Testing, Parel, Mumbai 400 012, INDIA

[3]Department of Microbiology, P.E.S. Modern College of Arts, Science and Commerce, Shivajinagar, Pune 411 005, INDIA [#]Quality Control Division, Serum Institute of India, Hadapsar, Pune 411 028, INDIA

## Abstract

**Spaceflights and clinostats have been used extensively to study the effects of microgravity on various biological systems ranging from microbes to plants. Similarly hypergravity studies have been carried out using centrifuges where growth retardation has been observed. However, no studies have been carried out yet on how the gravity of astronomical bodies, e.g. Moon having $1/6^{th}$ the gravity of Earth, affects biological systems. Such studies are important with missions to Moon and Mars to be**

*Biophysics Laboratory, Department of Physics, University of Pune, Pune 411 007, INDIA

Fax: +91-44-22232711

Email: abes137@gmail.com

Present Address

[@]Life Sciences Division, AU-KBC Research Centre, MIT Campus, Chennai 600 044, INDIA

carried out in future. Also, a comparative study to see the effects of gravity that exists on astronomical bodies such as Moon, Mars and Jupiter on any organism using simulation have not been reported so far. This paper discusses the effects of modelled gravity on the growth of *Vibrio harveyi* using the clinostat-centrifuge system designed and developed in-house. Results showed that though growth as measured by optical density was significantly higher for simulated microgravity and lunar and Martian gravities, there was no significant difference in viable counts. This is because the relative death rate is also higher for these gravities. Jovian gravity was found to slightly retard the growth. This study also shows that simulated lunar gravity is relatively most suited for the growth of *Vibrio harveyi*.

**Running Title: Growth of *Vibrio harveyi* in altered gravity**

**Introduction**

Experiments on *Escherichia coli* in space showed a shortened lag phase, an increased duration of exponential growth and an approximate doubling of final cell population density compared to controls (Klaus *et al.*, 1997; Gasset *et al.*, 1994) but no change in growth rate was observed (Brown *et al.*, 2002). Modelled microgravity also showed an enhanced growth of *E. Coli* while hypergravity retarded the growth (Gasset *et al*., 1994). *Bacillus subtilis* showed a higher growth rate in microgravity (Mennigmann & Lange, 1986; Kacena *et al*., 1999) as well as higher final biomass yield (Mennigmann & Lange, 1986). *Bacillus subtilis* also showed a shortened lag phase during spaceflight at 23 °C but not at 37 °C (Kacena *et al*., 1999). *Salmonella typhimurium* showed a higher growth rate and reduction in generation time under modelled microgravity (Wilson *et al*., 2002). An increased growth was observed in spaceflight as well (Mattoni, 1968). Modelled microgravity decreased the germination efficiency of *Dictyostelium discoideum* while hypergravity promoted it (Kawasaki *et al*., 1990). Fruiting bodies were found to be smaller in modelled microgravity and taller in hypergravity (Kawasaki *et al*., 1990). Paramecium showed a random swimming behaviour below 0.16g with negative gravitaxis becoming pronounced from 0.3 g suggesting a threshold for gravitaxis between 0.16 and 0.3 g (Hemmersbach et al, 1996). Studies on the contraction activity of Physarum onboard SPACELAB-I showed that the threshold was 0.1 g to elicit a response (Block et al, 1996).

In this paper, we discuss how modelled gravity conditions of Moon (0.16g), Mars (0.38g) and Jupiter (2.5g) affects the growth of *Vibrio harveyi*, a bioluminescent bacterium, isolated from the coastal waters of Goa, India. Bioluminescent bacteria have applications as environmental biosensors and pollution indicators (Girotti *et al*., 2008). We show that though microgravity, lunar and Martian gravities enhance the growth, they also enhance the death of *Vibrio harveyi*. To the best of our knowledge, neither studies on the effects of lunar and Martian gravities (hypogravity) nor a comparative study of different gravity environments have been carried out on the growth of any organism.

## Materials and Methods

Bacterial cultures obtained from coastal waters of Goa, India were isolated and identified as *Vibrio harveyi* (strain NB0903) using 16S rRNA sequencing. The suspension culture was adjusted to OD = 0.2 (approx. $5 \times 10^9$ cells/ml) and 1% inoculum added to sterile BOSS broth (Klein *et al.*, 1998) in each of four identical glass vessels (8 cm x 11 cm), specially designed for the clinostat-centrifuge system (designed and developed in-house) (Fig.1). Two of these cultures were exposed to modelled varied gravity conditions viz. Microgravity (Space), 0.16 g (Lunar), 0.38 g (Martian), 1 g (Earth) as

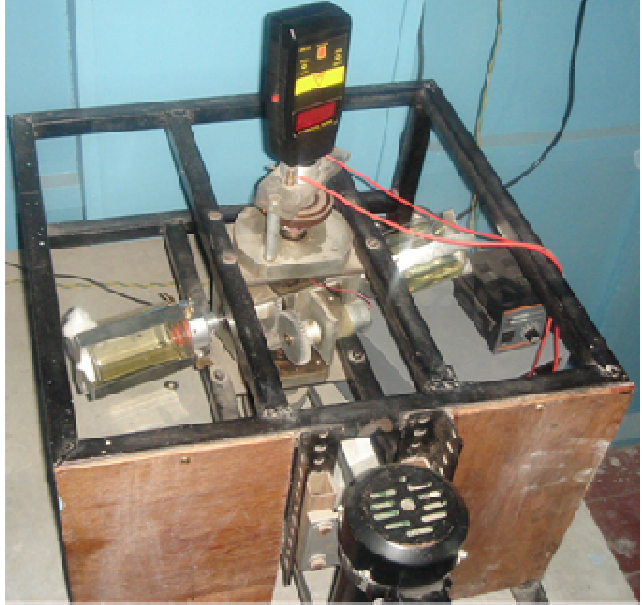

Fig 1: Experimental Setup

dynamic control and 2.5 g (Jovian) for 15 hours. The g-values were calculated using the standard formula

$$RCF = 1.118 \times 10^{-5} \times r \times N^2 \quad \ldots \quad (1)$$

where

RCF is the relative centrifugal force in g units,

R (=17 cm) is the distance of the sample from the centre of the clinostat-centrifuge in cm. This includes the distance of the sample holder from the centre and half the length of the vessel as the rotation is along the vertical axis. In case of microgravity r (= 4 cm) is the radius of the vessel as the rotation is along the horizontal axis.

and

N is the speed of rotation in rpm

In a normal centrifuge, two forces viz., gravitational force downward and the centrifugal force outwards, act on the sample. The net force acting on the sample is the resultant of the two and hence the magnitude of the resultant would be always higher than 1g. However in a clinostat, the sample is in modelled microgravity. Thus in a clinostat-centrifuge system, accelerations less than 1g can also be modelled.

The other two cultures acted as static controls. Static controls were used since shaking at 170 rpm would be equivalent to 3g. From the time of inoculation, at regular intervals of one hour, aliquots of 2 ml were taken for measuring optical density at 590 nm using a digital colorimeter and for total viable count using spread plate method in duplicates. Both control and test vessels were shaken well before taking the aliquots so that the cultures were well-mixed. Changes, if any, in morphological characteristics were also observed.

Since OD is a function of the total cell number for a particular instant, the difference in OD and TVC can be correlated to the number of dead cells at that instant. The relative death rates for different values of g were calculated as follows. Log CFU versus OD in the exponential phase was plotted and a mathematical function relating them was obtained. Taking this relation as a standard, the total number of cells in the broth for each value of g was estimated by substituting the corresponding OD values in this relation. The difference between the total number of cells in broth and CFU gives us the relative number of dead cells. This was used to obtain the relative death rate. Since the relation for static control is taken as zero, the relative death rate would be zero.

All experiments were carried out at room temperature ($25\pm1$ °C). Each experiment was repeated three times and consistent results were obtained. All data are represented as Mean ± SEM. The p-values were obtained by Student's T-test.

## Results and Discussion

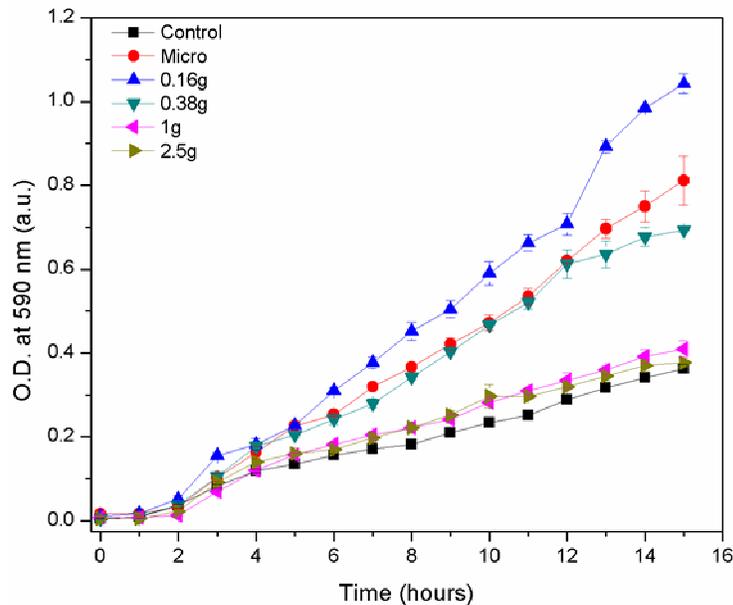

Fig. 2. Growth Curves by Optical Density

Significant differences were observed between the growth curves (by OD at 590 nm) under modelled microgravity ($p < 0.0002$), lunar ($p < 0.0002$) and Martian ($p < 0.005$) gravity conditions and those of control for all time points after the exponential phase began viz., 3 hours (Fig. 2). Comparison of OD showed maximum growth in modelled lunar gravity followed by modelled microgravity compared to static control. This was followed by modelled Martian gravity, Earth's gravity/ dynamic control and Jovian gravity compared to static control. Our results for modelled microgravity are in agreement for those obtained for *E. Coli* (Brown *et al*., 2002, Kacena *et al*., 1999, Klaus *et al*., 1997) and *B. Subtilis* (Mennigmann & Lange, 1986, Kacena *et al*., 1999) under spaceflight and clinorotation conditions.

However the viable cell counts for each value of g did not show any significant difference with respect to static control (Fig. 3). Space-flown bacteria grown on agar plates have shown similar results (Kacena *et al*., 1997). Similar results were obtained for *Pseudomonas aeruginosa* grown under modelled microgravity (Guadarrama *et al*., 2005). Some studies have attributed this to the influence of motility (Thevenet *et al*., 1996, Benoit & Klaus, 2007) while some others have attributed this to lack of more efficient metabolic capabilities in microgravity (Kacena *et al*., 1997). It should however be noted that in our case, liquid cultures were grown under altered gravity while viable counts for all the treated cultures were obtained under normal gravity on agar plates using the aliquots taken from these liquid cultures. Yet similar results have been obtained.

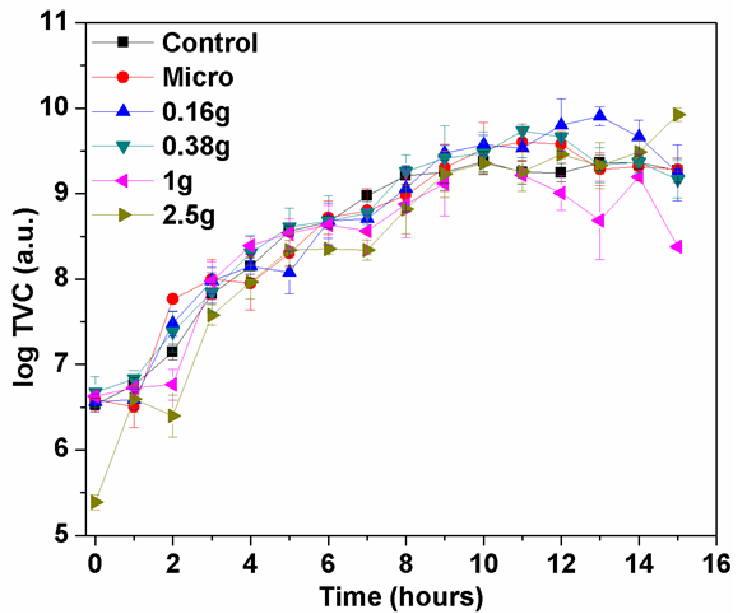

Fig. 3. Growth Curves by Total Viable Count

From Table 1, it can be seen that growth rate from OD was the highest for modelled lunar gravity. This was followed by the growth rate for modelled microgravity and modelled Martian gravity which were almost the same while Jovian gravity showed the minimum growth rate. However, the relative death rate is also higher in modelled microgravity and lunar and Martian gravities. This shows why the viable cell counts are not significantly different from control. Interestingly, the relative death rate is lesser than the growth rate from OD only for lunar gravity. This is reflected in the form of least generation time and maximum growth rate from viable cell counts. This shows that in effect; modelled microgravity, lunar and Martian gravities enhance the growth of *Vibrio harveyi* with modelled lunar gravity showing maximum enhancement while Jovian gravity retards the growth. The lag phase was reduced by 50 % for modelled microgravity, lunar and Martian gravities but remained the same for Jovian gravity with respect to control. As far as the exponential phase is concerned, it increased by 50 % in modelled microgravity, 33 % in modelled lunar gravity and 17 % in modelled Martian gravity but remained the same in Jovian gravity with respect to control.

No changes in colony morphological characteristics were observed for all samples exposed to each of the altered gravity conditions. SDS-PAGE did not show any change in band patterns for all samples (data not shown).

We expected an inverse relation between the growth and gravity and indeed it is the case except for lunar gravity. Hence it was quite surprising to get maximum growth for lunar gravity instead of microgravity. These results can be concluded due to gravity alone and not due to other factors such as aeration since growth for Jovian is less than 1g control. If it was the case, then maximum growth should have been observed for Jovian gravity since the rotation speed is maximum and hence the aeration relative to all other values of g reported in this paper.

In summary, we have measured the growth rates by OD and viable counts under different modelled gravity conditions, however a significant difference is found only in the OD. Organism under modelled microgravity, lunar and Martian gravity conditions shows, higher growth rate and reduced generation time with respect to static controls while Jovian gravity did not have a significant impact on the organism studied. Non-significant difference in viable counts is due to an equal number of cells dying under these conditions and not possibly due to similar metabolic activities (Kacena *et al.*, 1997) in microgravity or differences in motility (Benoit & Klaus, 2007).

We have shown that modelled lunar gravity, i.e. 0.16g, is most favourable for the growth of *Vibrio harveyi*. However, it is important to know if our results are a general trend for all organisms or are typical of *Vibrio harveyi* alone. Similar studies on mammalian cells are important, since they will provide an insight as to how our physiological systems will adapt to such gravity conditions. This has great implications for setting up bases on other planets and their moons. This can also be used in industry to get maximum output from micro-organisms just by optimising the value of g in addition to pH, temperature, etc.

## Acknowledgements

We are thankful to the Director, National Centre for Cell Science, Pune, India as well as to Prof. B.A. Chopade and Mr. .Praveen K. Sahu, DNA sequencing Laboratory, Institute of Bioinformatics & Biotechnology, University of Pune, India for carrying out the 16S rrna sequencing of the bacterial isolate. Comments provided by Mr. Vivek Jadhav, University of Pune, Mr. Anant Rajeha, I.I.T. Chennai, Mr. Lasse Folkerson, Karolinska Institute, Stockholm and Ms. Jamila Siamwala, AU-KBC Research Centre, Chennai were of great help in writing this paper. Co-author PBV would like to thank abdus-Salam International Centre of Theoretical Physics, Trieste for Associateship.

**Author Contributions** SB, AS and RD conceived the idea of this work. AS and SB made designs of the vessels while RD and JD isolated and characterised the organism. SB and AS designed the experiments. RD, JD, AS and SB contributed equally to the growth curve experiments. JD carried out the SDS PAGE.  SB and JD mainly wrote the paper. PBV supervised this work.

**Table 1. Growth kinetics in altered gravity**

| g | Growth Rate from OD (Hour$^{-1}$) | Relative Death Rate (Hour$^{-1}$) | Growth Rate from TVC (Hour$^{-1}$) | Generation Time (Minutes) | Lag Phase (Hours) | Exponential Phase (Hours) |
|---|---|---|---|---|---|---|
| Static Control (1g) | 0.365 ± 0.038 | - | 0.365 ± 0.038 | 114.9 ± 11.12 | 2 | 6 |
| Space (μg) | 0.51 ± 0.043 | 0.651 ± 0.174 | 0.392 ± 0.043 | 107.01 ± 11.75 | 1 | 9 |
| Moon (0.16g) | 0.602 ± 0.044 | 0.57 ± 0.232 | 0.41 ± 0.056 | 102.65 ± 12.56 | 1 | 8 |
| Mars (0.38g) | 0.432 ± 0.042 | 0.485 ± 0.171 | 0.394 ± 0.037 | 106.38 ± 9.65 | 1 | 7 |
| Dynamic Control (1g) | 0.294 ± 0.042 | 0.436 ± 0.159 | 0.383 ± 0.105 | 114.92 ± 12.83 | 2 | 6 |
| Jupiter (2.5g) | 0.235 ± 0.041 | 0.258 ± 0.128 | 0.34 ± 0.037 | 123.26 ± 12.89 | 2 | 6 |